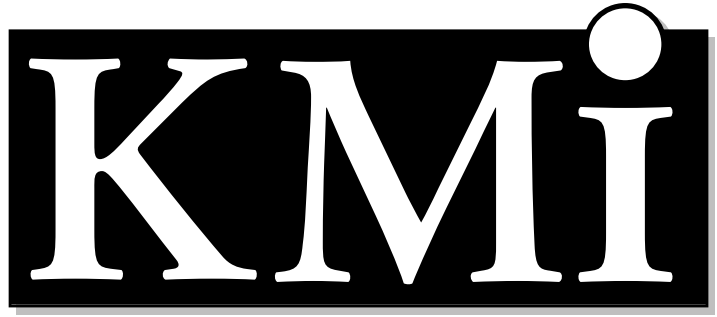

# Knowledge Media Institute

# Representing Scholarly Claims in Internet Digital Libraries: A Knowledge Modelling Approach

*Simon Buckingham Shum, Enrico Motta and John Domingue*

**KMI-TR-80**

**July, 1999**

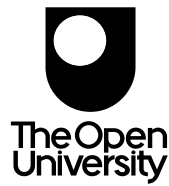

# Representing Scholarly Claims in Internet Digital Libraries: A Knowledge Modelling Approach

Simon Buckingham Shum, Enrico Motta and John Domingue

Knowledge Media Institute, The Open University, Milton Keynes, MK7 6AA, U.K.
{S.Buckingham.Shum, E.Motta, J.B.Domingue}@open.ac.uk
http://kmi.open.ac.uk/projects/scholonto/

**Abstract.** This paper is concerned with tracking and interpreting scholarly documents in distributed research communities. We argue that current approaches to document description, and current technological infrastructures particularly over the World Wide Web, provide poor support for these tasks. We describe the design of a digital library server which will enable authors to submit a summary of the contributions they claim their documents makes, and its relations to the literature. We describe a knowledge-based Web environment to support the emergence of such a community-constructed semantic hypertext, and the services it could provide to assist the interpretation of an idea or document in the context of its literature. The discussion considers in detail how the approach addresses usability issues associated with knowledge structuring environments.

## 1 Introduction

The mere publication of information does not constitute a body of knowledge; nor does simply obtaining information constitute understanding. Obtaining documents is just the first step; meaning and significance arise through their *interpretation*, which results in an understanding of the *perspective* adopted. In this paper we describe a representation scheme implemented in a Web server architecture to assist scholars in articulating, interpreting and contesting perspectives. Specifically, these perspectives take the form of networks of claims about ideas and documents. We propose that researchers enrich their texts with nodes and links which they add to a semantic network.

The paper begins by considering tasks that face scholars in analysing a document or literature. We argue that current approaches to document description, and current technological infrastructures, particularly over the World Wide Web, provide poor support for these interpretive tasks. We describe an approach to modelling the perspective in which documents are embedded, based on researchers' claims about their own work, and those of others. We detail how such a model is being implemented in a Web environment, and consider the services it could provide to a scholarly community. These include the creation, visualization and interpretation of conceptual structures reflecting the relations between different research efforts, scholarly perspectives and debates, which we exemplify with a worked example. We then discuss related work, key issues that this work raises, and the next steps in our research programme.

## 2   The scholarly work of interpretation

*Contextualising* ideas in relation to the literature is a fundamental task for authors and readers—are they new, significant, and trustworthy? Scholars usually start by bringing to bear their own knowledge of the field. This often leads to *commentary and discourse* of various kinds, which reflect the extent to which peers regard an author's work as authoritative, ranging from private annotation of a document, to formal peer review of conference/journal submissions, to published reviews of literatures and books. In the context of annotation and peer review, we have described elsewhere a publishing toolkit [25] that converts a scholarly document into a structured discussion website, and an electronic journal [13] which uses this to support an innovative peer review model.

We can think of conventional scholarly publication and debate as a document-centred, text-based process. Text is a rich medium in which to publish and discuss ideas in detail and with subtle nuances, but the corresponding disadvantage is that it takes a long time to read, and is hard to analyse computationally. We have demonstrated in previous work that a document-centric use of the Web can transform peer review in important ways [9], [39]. The complementary approach described in this paper, with different goals, focuses on the conceptual models implicit in textual documents and discourse. The goal is to provide a *summary* representation of ideas and their interconnections, in order to assist literature-wide analysis. We propose that this has advantages over textual media for tracing the intellectual lineage of a document's ideas, and for assessing the subsequent impact of those ideas, that is, how they have been challenged, supported and appropriated by others. In addition, the availability of explicit conceptual models opens possibilities for automatic assistance in analysing a community's (published) collective understanding of ideas.

We begin with the idea that an author's goal is to persuade the reader to accept his/her perspective, which constitutes a set of claims about the world. Usually, the author has some new ideas that s/he is contributing, and asserts particular relationships between these and existing ideas already published in order to demonstrate both the reliability of the conceptual foundation on which s/he is building, and the innovation and significance of the new ideas. The reader's task is to understand which ideas are being claimed as new, and assess their significance and reliability. Moving to the task of literature search and analysis, in this case, the scholar has some ideas and relationships in mind that s/he is trying to locate in the literature—has anyone written about them, or perhaps these ideas exist but not yet in a single document? The interpretive task includes formulating the ideas of interest in a variety of ways that may uncover relevant documents, reading the documents and then interpreting them to characterise any patterns that appear to emerge. This is a similar scenario to that of a newcomer to a scholarly community (e.g. a student; librarian; lecturer or researcher from another discipline) who wants to know, for instance, what the seminal papers are, or if there are distinctive perspectives on problems or classes of technique that define that community.

Scholars are poorly supported in these tasks by conventional library environments, physical or digital. Consider the document interpretation task. In the non-digital world, there is currently no way beyond following citations (only those provided by the author), or using citation indices (to find others citing him for some reason), to ask questions such as:

- Has anyone built on the ideas in this paper, and in what way?
- Has anyone challenged this paper?
- Has anyone proposed a similar solution but from a different theoretical perspective?

Considering the literature analysis task, there is currently no way for a scholar to query a digital library with analytical/interpretive questions of the following sorts:

- Are there any documents building on theory T, but which contradict each other's predictions?
- Are there any documents applying method M to domains D and E?
- Is there any Web-based software which tackles problem P?
- What impact did Theory T have?

- Are there distinctive theoretical perspectives on problem P?

These are the kinds of phenomena of most interest to scholars when they write papers, engage in debate or search the literature. These are also the kinds of questions asked by researchers unfamiliar with a literature, including students. Our approach seeks to provide better support for identifying significant conceptual structures that a research community considers important.

## 3  Evolving the Web beyond simple linking

The World Wide Web is the first global hypertext system to emerge, providing a rudimentary infrastructure for publishing interlinked documents and discourses. This level of representational sophistication is already extremely useful for enhancing, even transforming, scholarly publication and discourse, as we have doumented elsewhere [9], [39]. However, the Web provides little support for *structuring*, *searching* or *analysing* scholarly concepts, documents or discourses. Early, pre-Web hypertext systems have already demonstrated (on a small scale) the power of features such as semantically typed nodes and links, bidirectional links, composite nodes which represent more complex structures, and structural searching. It is increasingly recognised that the Web would benefit from such features (see for instance an analysis of 'fourth generation' Web functionality [4]).

Information retrieval using statistical and text analysis techniques include techniques for clustering and mapping documents based on semantic similarity [11], [12], and for inferring certain types of inter-document relationships (e.g. "cites", or "summarises") [1]. Automatic techniques clearly have the advantage that they can be applied to large text corpora with little human effort once the texts are in a suitable format for processing. From the perspective of scholarly interpretation, the key weakness of such techniques is that it is extremely hard, if not impossible, to automatically identify more complex kinds of *scholarly relationships* between documents such as those given above. *Human-encoded* document descriptions are required to express such scholarly claims and structures.

We propose that when a new article is ready for dissemination, authors describe the document's main contributions and relationships to the literature using a controlled vocabulary analogous to a metadata scheme (but implemented using a formal *ontology*), and submit the description to a networked repository. We describe in the next section the concept of an ontology, and the representational scheme underpinning our approach.

## 4  Representing scholarly claims

### 4.1  Ontologies

An ontology in philosophy refers to a model of what exists in the world [10]. The artificial intelligence community has appropriated the term to mean the construction of *knowledge models* [18], [32] which specify concepts or objects, their attributes, and inter-relationships. A knowledge model is a specification of a domain, or problem solving behavior, which abstracts from implementation-centered considerations and focuses instead on the concepts, relations and reasoning steps characterizing the phenomenon under investigation. Our application of knowledge modelling in this project is to implement a semantic network which expresses important aspects of the web of ideas and perspectives implicit in the documents and minds of a scholarly community.

### 4.2  An ontology for representing scholarly claims

An ontology reflects a (typically community-wide) viewpoint on how best to conceptualise a particular domain or phenomenon. Hence, its main role is to support knowledge sharing and reuse. It might appear paradoxical, therefore, to propose the use of ontologies to support scholarly communities in managing their knowledge, since conflicting worldviews, evidence and frames of reference lie at the heart of research and debate.

The key issue is in what is represented. Our approach builds on a relatively stable dimension of what are otherwise constantly evolving research fields, by representing scholars' *claims about the significance* of ideas and concepts—a focus on *discourse* and *argumentation* (how scholars support and contest claims), and on *context* (the conceptual network in which an idea is embedded). In other words, it is hard to envisage when researchers will no longer need to make claims about, or contest, the nature of a document's *contributions* (e.g. "this is a new theory, model, notation, software, evidence"), or its *relationships* to other documents (e.g. "it applies, modifies, predicts, refutes…"). Moreover, separating the representation of concepts from claims about them will be critical to supporting multiple perspectives.

We are adopting a philosophy of 'minimal ontological commitment' [19] and incremental formalisation [37], which reflects an emphasis on making explicit just enough structure to be usefully expressive and enable the provision of valuable computational services, but leaving the *document texts* to express the details and nuances of an author's argument (as opposed to trying to formalise it). This minimises the effort required to submit a document description; if there is evidence that authors wish to link ontological concepts to specific paragraphs within a document (e.g. as proposed by [23]), then we can provide a way to do so, but we will begin at the document level. The kind of core scheme we are moving towards, suitable for a wide range of disciplines, is proposed in Figure 1, but it is both generalisable and tailorable to other fields (e.g. a more experimental field might specialise *Idea* into *Hypothesis*; another might not need *Software*).

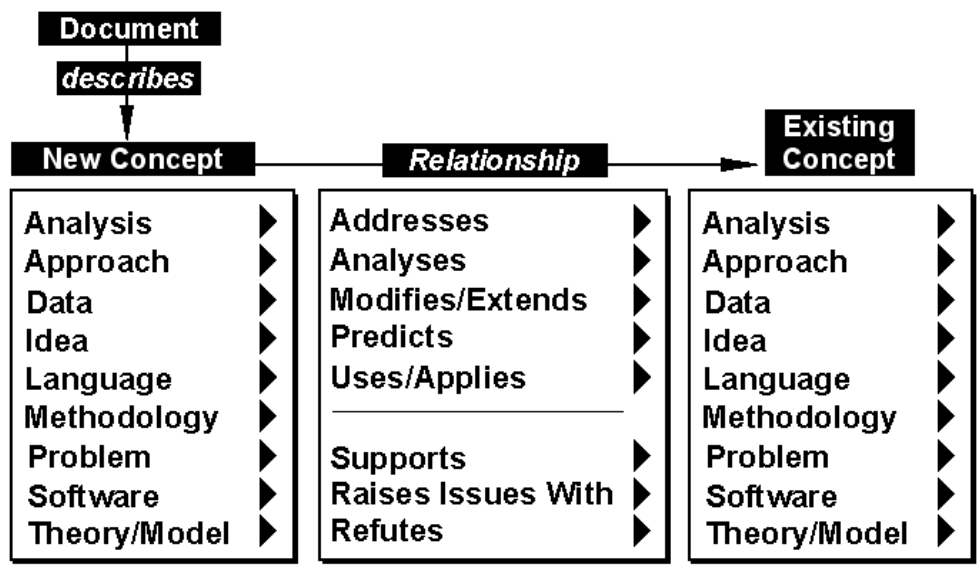

**Fig. 1.** Representational elements for summarising the key contributions of a publication, and their relationships to other concepts.

We hypothesise that a research community should be able to agree on a relatively small set of uncontroversial conceptual and relational types which can adequately express the majority of claims made. The goal is to design an ontology which is simple enough to understand without being simplistic, yet expressive enough that most researchers can represent the key claims made in most documents.

More elaborate argumentation schemes have been proposed for computer-supported argumentation (e.g. [26], [34], [38], [40]), but our analysis of this literature shows little evidence of successful uptake (see discussion). The ScholOnto scheme therefore supports argumentation in relatively simple terms (*supports, raises issues with,* and *refutes*) to make it as easy as possible to add an argumentation link to a concept or document; more elaborate schemes can be introduced when there is the demand from a community. Moreover, whilst rigorous and carefully maintained argumentation networks make many kinds of useful analysis possible (the main motivation for increasing a schema's expressiveness), such tools make assumptions about users' expertise and consistency of representation that are unlikely to hold in the context of an open, internet community. The details of an author's reasoning are therefore left to the document's text, and are not made explicit in the knowledge base.

To summarise, we propose that this represents a novel approach to the persistent problem facing any ontology development effort, namely, that the world being described is typically dynamic, necessitating resource intensive updating and restructuring. Shifting the representational focus to the way in which researchers make new contributions to the literature avoids the problem of committing to a domain ontology that quickly becomes outdated. The domain ontology is only constructed in the context of authors' claims about their work, and can be contested by others.

## 5 Implementation

### 5.1 Knowledge modelling infrastructure

Our approach relies on a suite of robust knowledge modelling technologies developed and tested in other domains. The *OCML* modelling language [30] supports the construction of knowledge models by means of several types of construct. It allows the specification and operationalization of functions, relations, classes, instances and rules. It also includes mechanisms for defining ontologies and problem solving methods [3], the main technologies developed in the knowledge modelling area. Problem solving methods are specifications of reusable problem solving behaviours. OCML has been used in several projects, in domains such as medicine, geology, engineering design and organizational learning. As a result the language is now associated with a large library of reusable models, providing a useful resource for the knowledge modelling community. In our scenario, OCML provides the formalism for defining our ontology for scholarly debate and interpretation, henceforth referred to as *ScholOnto*.

*WebOnto* [14] enables knowledge engineers to collaboratively browse and edit knowledge models over the Web. The architecture is composed of a central server and a Java applet. WebOnto's central server is built on top of a customised web server *LispWeb* [33] and uses OCML as the underlying modelling language. In addition to implementing the standard HTTP protocol, the LispWeb server offers a library of high-level Lisp functions to dynamically generate HTML pages, a facility for dynamically creating image maps, and a server-to-server communication method. The WebOnto Java applet provides multiple visualizations of OCML knowledge models, a direct manipulation and forms interface for creating new knowledge structures, and a groupware facility which supports both synchronous and asynchronous model building by teams of knowledge engineers (illustrated in Figures 2 and 9). Applied to the problem of managing scholarly knowledge concepts and documents in online research communities, these technologies provide the building blocks for a scaleable Web infrastructure.

Further details on these tools and our approach to enriching documents with ontologies can be found in [31].

### 5.2 Ontology design

Figure 2 shows the top level structure of the ontology, as specified in OCML. Both nodes and links in the semantic network created by scholars' submissions are SCHOLARLY-KNOWLEDGE-CONCEPTs. Nodes are SCHOLARLY-CONTRIBUTION-ELEMENTs, and links SCHOLARLY-RELATIONSHIPs, which are subdivided into ARGUMENTATION-LINKs and NON-ARGUMENTATION-LINKs.

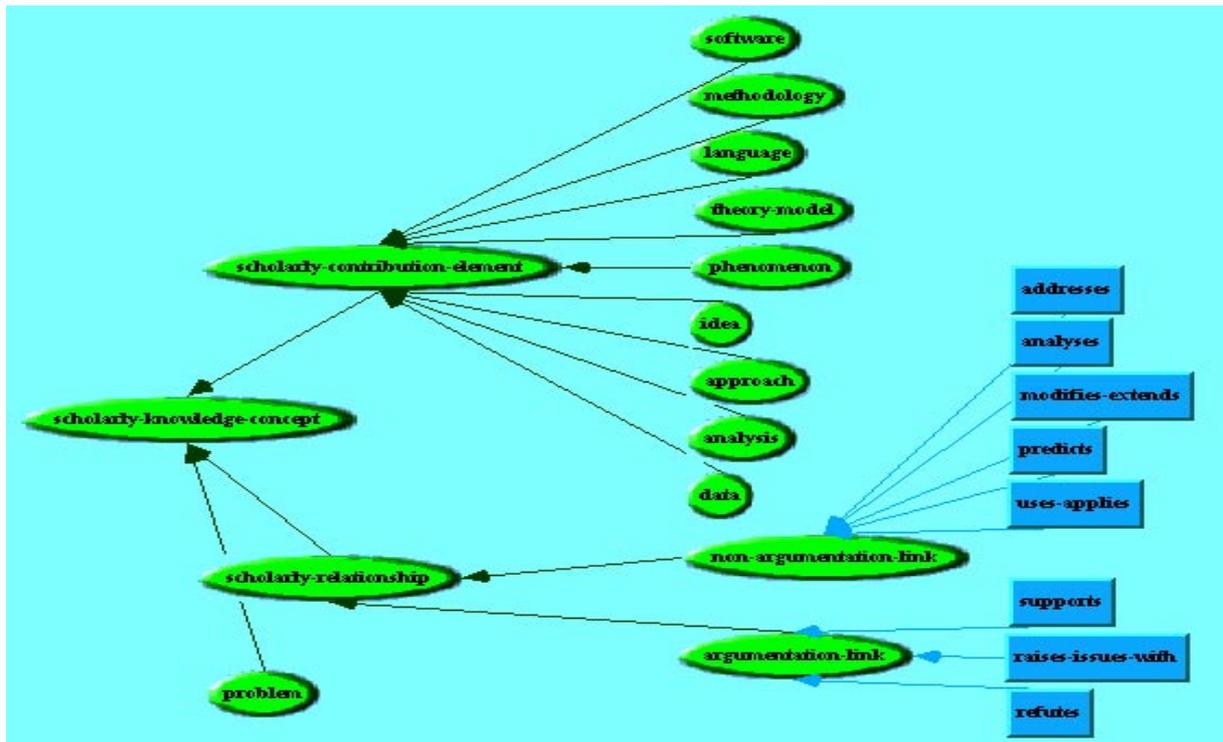

**Fig. 2.** Class structure of the *ScholOnto* ontology, represented in the *WebOnto* system.

Figure 3 shows the class definitions for the scholarly concepts of SOFTWARE, METHODOLOGY and LANGUAGE. SOFTWARE is defined as a SCHOLARLY-CONTRIBUTION-ELEMENT, which ADDRESSES PROBLEMS, USES/APPLIES any other kind of SCHOLARLY-CONTRIBUTION-ELEMENT (e.g. a METHOD or LANGUAGE), and MODIFIES/EXTENDS other kinds of SOFTWARE. METHODOLOGY and LANGUAGE are similarly defined.

```
(def-class SOFTWARE (scholarly-contribution-element)
 ((addresses :type problem)
  (uses-applies :type scholarly-contribution-element)
  (modifies-extends :type software)))

(def-class METHODOLOGY (scholarly-contribution-element)
 ((addresses :type problem)
  (modifies-extends :type methodology)
  (uses-applies :type scholarly-contribution-element)))

(def-class LANGUAGE (scholarly-contribution-element)
 ((addresses :type problem)
  (uses-applies :type language :type theory-model)
  (modifies-extends :type language)))
```

**Fig. 3.** OCML definitions of SOFTWARE, METHODOLOGY and LANGUAGE.

The ontology is designed to support scholars in making *claims* by asserting relationships between concepts. Other scholars may support, raise-issues-with, or refute these claims. Figure 4 shows schematically the structure of a scholarly "claim" in the ontology. The OCML specification associated with the structure in Figure 4 is shown in Figure 5.

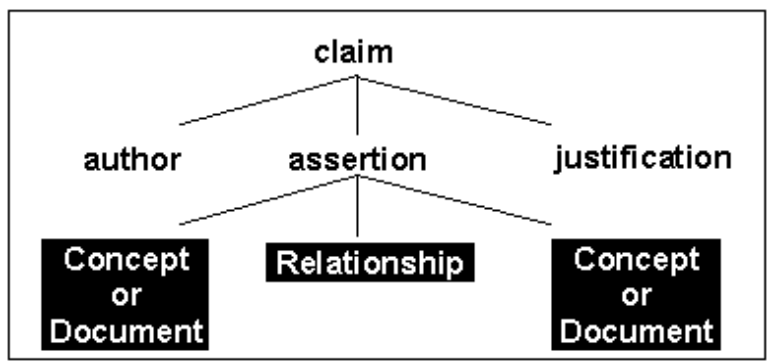

**Fig. 4.** The structure of a scholarly "claim" in the ontology.

A `claim` is formally defined as a relation between a set of `authors`, who make a `legal-scholarly-assertion`, with some `justification`. A `legal-scholarly-assertion` is a statement instantiating a `scholarly-relationship` (e.g. `addresses, predicts, refutes`) between two elements (e.g. `methodology X addresses problem Y`). The `justification` is free text supporting a claim. An author will not be expected to enter this since the justification for their claim is already to be found in the document they are describing. But if another scholar, for example, `supports`, `raises-issues-with`, or `refutes` another's claim (see scheme in Figure 1), without publishing an associated document, then some form of textual justification is expected. This could in turn point to a more rigorous justification in another document (ideally, directly accessible).

```
(def-relation claims (?X ?Y ?Z)
  :constraint (and (set ?X) (every ?x author)
                   (legal-scholarly-assertion ?Y)
                   (justification ?Z)))

(def-class legal-scholarly-assertion (assertion) ?x
  :iff-def (and (assertion ?x)
                (== ?x (?a ?b ?c))
                (scholarly-relationship ?a)))

(def-class justification (string))
```

**Fig. 5.** Separating claims from scholarly relationships; there can be many different (possibly contradictory) claims by different authors.

To both `support` and `refute` a particular claim is usually a sign of inconsistency, or perhaps, of a position that has changed from an earlier paper. OCML's envionrment makes it easy to construct rules that could, for instance, check for instances where an author is a member of two sets of authors who have made conflicting assertions (Figure 6).

```
(def-relation inconsistent-position (?auth ?assertion)
  :constraint (and (author ?auth)
                   (legal-scholarly-assertion ?assertion)))
  :sufficient (and (member ?auth ?x)(supports ?x ?y ?z)
                   (member ?auth ?x2)(refutes ?x2 ?y ?z2)))
```

**Fig. 6.** An OCML rule for an agent to check for positions that both `support` and `refute` a particular claim. This might reflect an `inconsistent-position` or at least, claims meriting closer examination.

Encapsulating such rules in agents that researchers could select from a library and tailor is a scenario that the ScholOnto architecture aims to support.

## 6 A Worked Example

We now return to our opening examples of literature and document interpretation, and use a worked example to clarify how our tools could support scholarly work. Within the hypertext research literature, one of the landmark articles is the summary of the *Dexter Hypertext Reference Model* by Halasz and Schwartz [22], which specifies both semiformally, and formally (using the Z notation), abstract properties of hypertext systems, enabling comparison of existing systems, and specification of theoretically possible future systems.

WebOnto already generates forms with contextual menus for adding new instances to an ontology (see [31]), but needs to be extended to generate forms that non-knowledge engineers can use. Figure 7 shows a prototype user interface for submitting the article's description to the repository. The user interface will guide users through the schema using dynamic menus, and enables them to browse and search for existing concepts to assist their reuse. Some domain concepts are simple to reference (e.g. the name of a specific software system, framework or methodology), whilst others are less concrete and could benefit from information retrieval support, e.g. finding the name(s) used to describe a domain problem ("user disorientation"), an idea ("a global hypertext system"), or an empirical phenomenon such as a piece of evidence ("low ability students benefit most from physics simulations").

This form would generate an OCML entry in the ontology, as shown in Figure 8.

```
(def-instance dexter-htxt-ref-model-article article
  ((describes-scholarly-contribution-element dexter-ht-ref-model
   (concerns-domain hypertext-hypermedia)
   (has-author halasz-f schwartz-m)
   (has-title "The Dexter Hypertext Reference Model")
   (publication-details "Communications of the ACM, 37 (2), 30-39")
   (has-url "www.acm.org/pubs/articles/journals/cacm/1994-37-2/p30-halasz/")
   (acm-ccs "I.7.2" "H.1.1" "H2.1" "H3.2" "H5.1")
))

(def-instance dexter-ht-ref-model theory-model
  ((addresses absence-of-standards
   (analyses notecards
             augment
             concordia
             hypercard
             hyperties
             intermedia
             kms-zog
             neptune-ham)
   (envisages theoretically-possible-dexter-compliant-systems)
   (uses-applies Z)))
```

**Fig. 8.** The OCML entry for the Dexter article, declaring its contributions to the literature (`dexter-ht-ref-model`, which is a `theory-model`, and `predicts` theoretically possible systems), and its relationship to other concepts (`analyses` several existing systems, and `uses-applies` the Z notation).

The article is now added to the *ScholOnto* knowledge base, enabling users to ask questions such as, *"What motivated the Dexter Hypertext Reference Model, and what impact has it had?"* A forms-based interface, generated automatically from the *ScholOnto* ontology by WebOnto, enables users to ask such questions through simple menu selection (Figure 9).

**Fig.7.** Prototype user interface for submitting claims about a document's contributions.

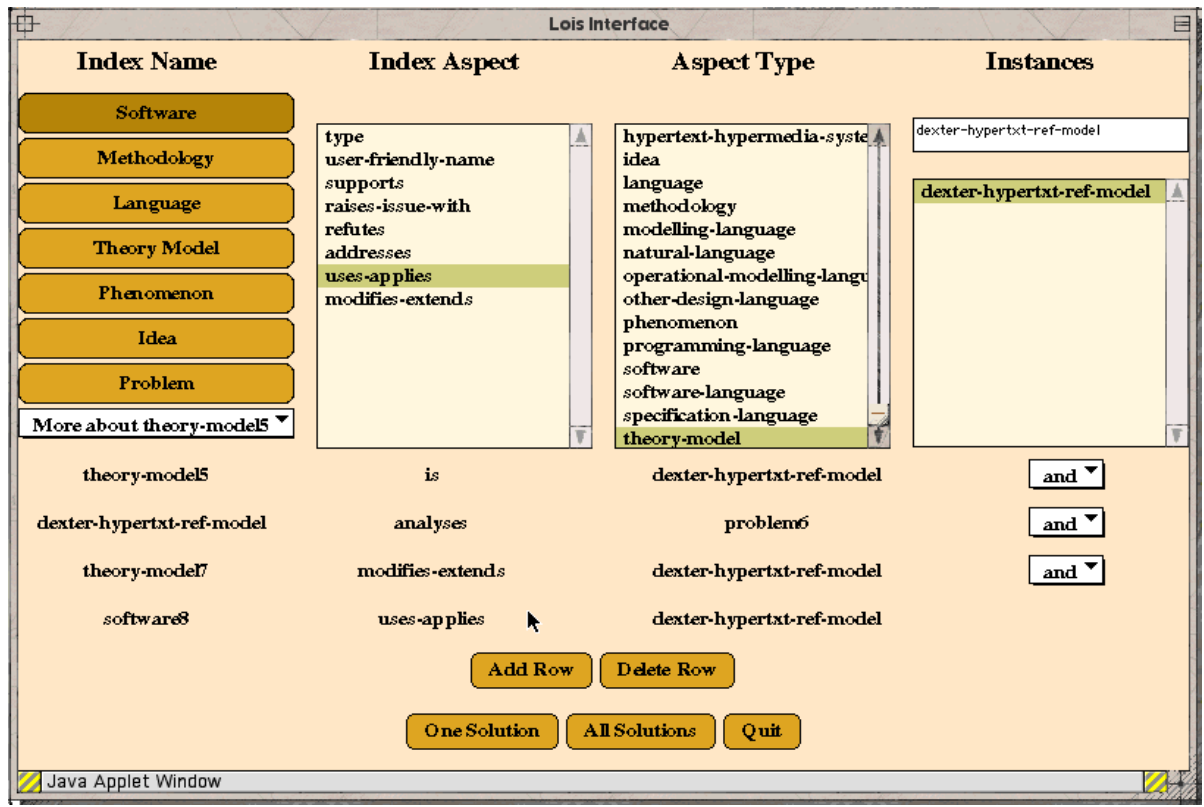

**Fig. 9.** A forms-based interface generated automatically from the *ScholOnto* ontology by WebOnto enables users to query the model via menu selection. The screenshot shows several possible queries to analyse the motivation behind, and impact of, the Dexter Hypertext Model (we have combined them to save space; in reality one would most likely submit these as separate queries). The queries specify, respectively, (1) interest in the `theory-model: dexter-hypertxt-ref-model`, (2) what `problems` does it `analyse?`, (3) are there any `theory-models` which `modify-extend` it?, and (4) is there any `software` which `uses-applies` it?

The knowledge base could generate filtered visualizations of the literature (e.g. based on the semantic network model in Figure 10) showing the Dexter Model's motivation and conceptual roots (links to the left), and the nature of the work which has built on it since by the respective authors, or other researchers (links to the right). This is an illustration of the kind of concept map that could be generated by a ScholOnto search.

## 7   Discussion

In this section we discuss some of the issues raised and possibilities opened up by the design we have presented, and contextualise our approach to related research.

### 7.1   Intelligent services

A knowledge model enables inference-based searching and alerting. It will be possible to ask the system questions such as "What impact did Theory T have?", since "impact" can be defined, for example, in terms of the number of subsequent documents using or modifying it, the number of different domains in which it has been applied, the number of problems addressed which drew on the theory, and so forth. Our knowledge modelling environment makes it simple for us (as system maintainers) to write heuristics that could assist in finding relevant documents, e.g. "if Method Y extends Method X, and Method X is challenged, then Method Y may be challenged". We will also assist scholars in composing their own rule-based interest profiles, e.g. "If 3 or more documents support Language L and challenge Language M (or any Languages based on them), and 3 or more documents support Language M and challenge Language L, then send me a concept map showing their interconnections"—since this may be evidence of two schools of thought.

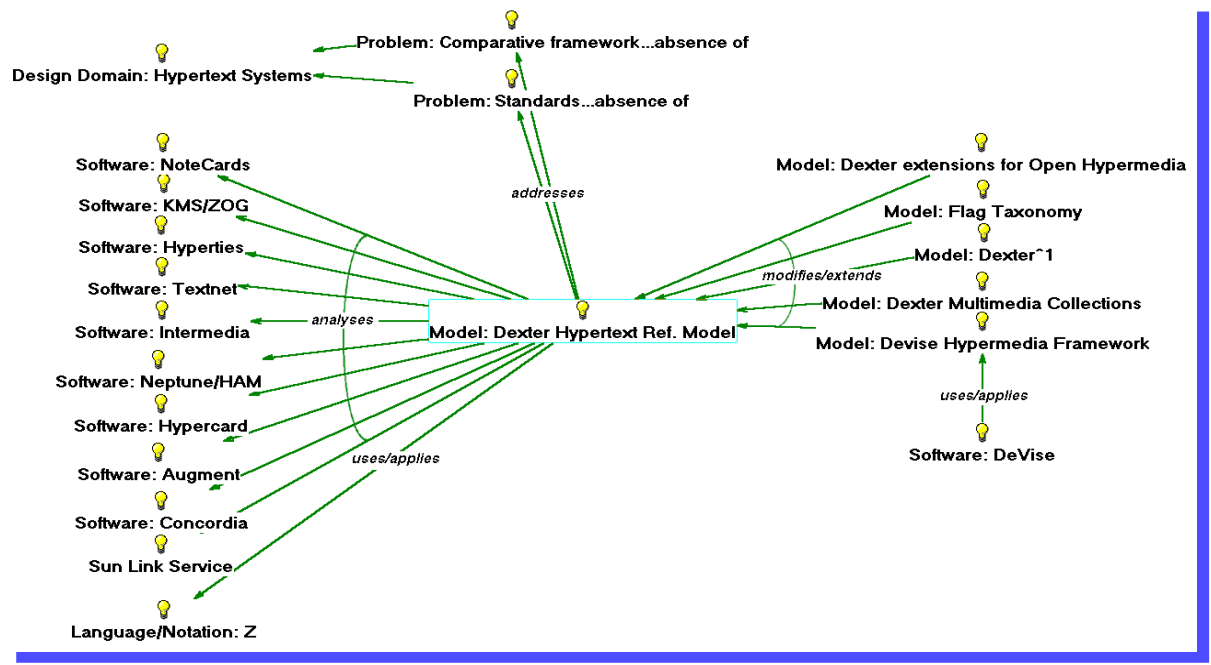

**Fig.10.** A semantic network model provides the basis for generating visualizations of the literature. In this example, an author has described the relationships (links to the left) that motivate and situate an article's key contribution (a Reference Model, marked by the central node) within its literature. Subsequent researchers (links to the right) have *modified/extended* the model, and implemented software systems based on their extended models (e.g. the "DeVise" system, lower right).

Another important advantage of our approach is that the existence of a formally represented knowledge model makes it possible to envisage additional reasoning services on top of the 'basic' search support. For instance, it will be possible to develop specialized agents whose goal is to identify emerging perspectives, using heuristic knowledge and machine learning techniques. For instance, an agent could discover a 'European perspective' on a particular issue, by analyzing the geographic spread of the relevant positions.

The complex notion of a 'school of thought' or theoretical perspective has representational form within ScholOnto. A perspective can be recognised by the common THEORY/MODELS to which a group of researchers appeals, the associated METHODS and LANGUAGES which they deploy, and the body of EVIDENCE that they mutually support. Conversely, the set of THEORY/MODELS, METHODS, LANGUAGES and EVIDENCE that they collectively RAISE-ISSUES-WITH may represent a different perspective. Tools to enable researchers to define such conceptual structures open intruiging new possibilities for literature tracking and analysis.

The 'impact' or 'authority' of a piece of work can also be represented in a variety of ways in ScholOnto, depending on author preference. Beyond quantitative counts of how many documents cite a document or reuse a concept (directly, or through more complex inheritance), one could also declare an interest in a particular methodology or theoretical perspective (see above) as carrying more weight, and filter the literature on this basis.

### 7.2 Usability of knowledge structuring tools

We are acutely aware that many schemes for registering shared resources and providing structured descriptions founder on the crucial 'capture bottleneck' – the envisaged beneficiaries of the system simply do not have the motivation or time to invest in sharing resources to reach a critical mass of useful material. We drew sobering lessons on this theme from an analysis of the computer-supported argumentation literature [7], corroborated by evidence relating to groupware [20], design rationale support [8], organizational memory systems [6], [35], and indeed, for many CSCW systems that require users to formalize information [36]. Specifically, we need to address issues raised by the usage patterns of hypertext systems such as Textnet [41] for scholarly annotation and linking, and descendants such as NoteCards [21] and Aquanet [28]. These provided rich schemas of semantic node-link types, but the limited adoption of these features led Trigg and others to conclude, correctly in our view, that rich taxonomies of node and link types (e.g. about 40 in Textnet) overwhelm users. However, before eradicating all human-encoded semantic hypertext links from our systems, we argue that effective use of such systems may depend on finding the right mix of target domain, context of use and user community.

Why should this proposal work where others have failed? Our target community, domain and use context have unique characteristics lacking in many other domains in which group memory, knowledge structuring and argumentation have not fitted well.

- *Research-oriented publishing.* Academic and other research publishing, in contrast to other genres, emphasises the kind of careful argumentation and analysis of domain structure required, and possibly fostered, by this approach. (In addition, for teaching purposes, students can be required by a course to analyse the conceptual networks for a given literature, and perhaps construct their own concept maps as part of an assignment.)
- *Strong motivation to disseminate work.* Making an impact through publications is a primary activity for academics. Having completed a new document, the author will want to maximise its 'digital presence' on the net by carefully encoding its contributions and connections to the existing literature.
- *Opportunity to reflect on how to represent ideas*. In contrast to the synchronous group working context in which many argumentation systems have been tested, scholars will be describing their documents in an individual setting, with time to reflect on how best to construct a network description.
- *A simple semantic schema complementing text.* We hypothesise that most researchers will have no trouble in understanding node and link types such as those in – they are the concepts of everyday research discourse. Nor are we requiring an author to make explicit a document's thesis at a fine granularity; the structural representation is a summary to assist the document's discovery, not a substitute.
- *Benefits deriving from Web scalability.* Previous research with pre-Web groupware and group memory systems has always focused on individual or small scale collaborative use. In a small team who already know each others' work, it is often hard to justify the overheads of information structuring in order to track documents and debates. This is in sharp contrast to the challenge of tracking and analysing developments in an international, evolving digital library. Using the Web as our collaboration and delivery medium has a second advantage: the size of a Web-based research community increases the chances of quickly building a critical mass of users, which will in turn improve the value of the services provided.

### 7.3 Researchers are not librarians

Internet-based digital libraries of the sort that concern this project will change the roles of librarian and scholarly researcher established by paper-based, geographically-based libraries. For an internet-based library to scale realistically, with potentially tens of submissions arriving every day, the only people who can be expected to initially describe new articles are the people with most motivation to maximise the visibility and impact of the work—the authors. However, authors are not librarians or knowledge engineers who traditionally have possessed the skills to do information classification. This raises two issues: whether scholars are able to describe their documents sufficiently well to enable the system to make use of their descriptions, and how to make the underlying description technologies accessible and understandable. We follow initiatives to develop metadata schemes for the Web (see next section) in assuming that given intuitive representation schemes and user interfaces, domain experts will be able to submit useful descriptions of their own work. These are, however, empirical questions that will be addressed through lab-based studies studying detailed interaction with the system, followed by broader field trials once the system is deployed in different research communities.

### 7.4 Emergent work practices

If a significant proportion of a research community adopted a digital library infrastructure of the sort proposed, such that it became a de facto standard to register new documents on the server, it is likely that it would effect a shift in working practices. In a more speculative mood, we briefly consider what forms these might take.

The public face of an authors' work would expand from his or her publications, to include the corresponding concept maps that others would see. Formulating the contributions of an article such that this map was succinct and to the point might in fact improve the authoring process. A research project or lab would most likely evolve a 'library' of conceptual elements and structures which they reused in different documents, and which represented their 'official' statement of how they wished their work to be linked to and reused by others.

A research field has a small number of recognisable 'genres' of article which an active member of that field can usually recognise very quickly. Indeed, these are formally recognised by some conferences and journals in order to guide reviewers on appropriate review criteria for different kinds of submission (systems paper, theory paper, empirical paper, etc.). We can envisage the possibility of a journal establishing a set of templates for submitting and reviewing submissions in different categories, that is, "a paper of type-X should reasonably demonstrate in its conceptual network representation its relationships to A, B and C, and we expect arguments of the form P, Q and R…". If the digital library is worth anything, it will of course assist reviewers in locating related work and debates that the author has omitted.

### 7.5 Relationships to other research

We have already addressed earlier hypertext and argumentation work; here, we situate our approach in relation to other relevant research and Web developments.

Our technologies can be seen as a conceptual and technical development from current efforts to develop metadata description schemes for the Web. Metadata in the context of digital libraries refers to ways to encode information about resources in machine-interpretable formats, typically by completing a standard set of descriptive fields. Well known examples include USMARC [27] for library resources, Dublin Core [15] to provide a simple high level scheme for web resources, and IMS [24] for educational resources. Our scheme for representing multiple claims about the status of content extends typical metadata schemes which normally focus on encoding uncontroversial content attributes. From a representational and technical perspective, our approach differs from metadata in that ontologies support more sophisticated modelling, for example, specifying sufficient and necessary conditions for relations, and providing metalevel modelling support which makes it possible to reason about the ontology itself. OCML in particular also provides powerful inference support making it possible to directly operationalise the ontology and its instantiation as a knowledge base. The W3C's proposed Resource Description Framework [42] for interrelating multiple metadata schemes employs semantic network modelling similar to the knowledge modelling concepts presented here, and could provide a future route to interoperability with other systems and metadata schemes.

The $(KA)^2$ initiative (Knowledge Annotation for Knowledge Acquisition) [2] aims to support the knowledge acquisition community in building a knowledge base of its own research by populating a shared ontology. The knowledge base is constructed by authors annotating their web pages (e.g.

publications; personal and project pages) with tags (analogous to HTML META tags), which can be read by a specialised search engine called Ontobroker [17]. The key architectural difference to our approach is that (KA)$^2$ semantic tags are embedded in the physical content, whereas our approach decouples content from claims about its status. This architectural difference reflects the different aims of the two enterprises. The aim of (KA)$^2$ is to capture the contents of web pages in a formalism which can be reasoned about by Ontobroker. In our scenario we do not aim to represent directly the domain-specific content of a paper, but the debate about the *scholarly status* of that content. Moreover, we are concerned that authors will not be prepared to invest the required effort to encode models of document content, and as argued earlier, we cannot assume a stable ontology to describe an active research field.

Ontologies are beginning to be used in the context of digital libraries, although for different purposes to those set out here. Ontologies can assist the extraction of concepts from unstructured textual documents [16], by serving as a source of knowledge about the particular topic. Ontologies can also assist in managing document descriptions in large digital libraries [43], [44].

Finally, our proposal builds on the use of automated pre-print servers for the submission and dissemination of documents (e.g. the Los Alamos server for physics and computer science <xxx.lanl.gov>, or the CogPrints server for cognitive science <cogprints.soton.ac.uk>). Hierarchical taxonomies and keywords assist search and e-mail alerting. A ScholOnto knowledge based server as proposed here extends this infrastructure by adding a semantically enriched layer of document encoding, with associated services, as discussed. One strategy for migrating from a conventional pre-print server to a ScholOnto server would be to import an existing pre-print server database, and enable authors to annotate their documents' entries to give them a presence in the semantic network.

## 8   Summary and Future Work

The Web has established itself as a medium for research document dissemination. However, its support for many of the interpretive tasks that scholars perform is weak, despite the fact that semantic hypertext systems (as the Web was originally envisaged by Berners-Lee) are well suited to tasks such as structural searching, pattern analysis, and heuristic filtering.

Our aim is to support scholarly analysis and discourse through the creation of author-centred and community-wide perspectives by researchers or software agents. We are developing a representational and technical infrastructure to complement libraries of archived documents with a 'living' semantic network of concepts. Adopted by a research community, this network could reflect the *evolving understanding and recontextualisation of ideas* over time (lost with archived documents), making possible new forms of literature analysis.

The proposed ontology expresses scholarly *claims* about domain concepts, not the domain concepts themselves; existing metadata and knowledge modelling efforts focus on describing the *content* of resources, rather than *discourse* about content. We focus on the conceptual models implicit in textual documents and discourse in order to provide a *summary* representation of ideas and their interconnections. This has advantages over textual media for tracing the intellectual lineage of a document's ideas, and for assessing the subsequent impact of those ideas. In addition, the availability of explicit conceptual models opens possibilities for automatic analysis of a community's collective knowledge.

We are currently finalising the core ontology to assist interoperability across a wide range of research domains, and seeding some test literatures to evaluate WebOnto's usability and automated services. We will then seek 'early adopter' communities interested in testing the ScholOnto server as a means of managing their own research knowledge.